\documentclass{Interspeech}

\usepackage{times}
\usepackage{latexsym}
\usepackage{booktabs}
\usepackage{verbatim}
\usepackage[table,xcdraw]{xcolor}
\usepackage{multirow}
\usepackage[table,xcdraw]{xcolor}
\usepackage{colortbl}  
\usepackage[table,xcdraw]{xcolor}
\usepackage{booktabs}

\usepackage{etoolbox}
\makeatletter
\patchcmd{\affiliation}
  {\renewcommand{\affiliationsep}{\vskip1pt}}
  {\renewcommand{\affiliationsep}{, }}
  {}{}
\makeatother

\usepackage[T1]{fontenc}
\usepackage[utf8]{inputenc}
\usepackage{microtype}
\definecolor{lightred}{rgb}{1.0, 0.8, 0.8}
\definecolor{lightblue}{rgb}{0.8, 0.9, 1.0}
\definecolor{lightgreen}{rgb}{0.8, 1.0, 0.8}
\definecolor{lightyellow}{rgb}{1.0, 1.0, 0.8}
\definecolor{lightpurple}{rgb}{0.9, 0.8, 1.0}
\definecolor{lightorange}{rgb}{1.0, 0.9, 0.8}
\usepackage{inconsolata}
\usepackage{adjustbox}
\usepackage{ subcaption, algorithm2e, url, float, etoolbox, adjustbox, pgf,booktabs, verbatim}
\usepackage{xcolor}
\usepackage{amsmath}
\usepackage{amssymb}
\usepackage{tcolorbox}
\definecolor{emerald5}{RGB}{236,247,238}   
\definecolor{emerald10}{RGB}{200,230,210}  
\definecolor{emerald15}{RGB}{164,214,185}

\title{Towards Source Attribution of Singing Voice Deepfake with Multimodal Foundation Models}

\author[affiliation={1}]{Orchid Chetia}{Phukan*}
\author[affiliation={1,2}]{Girish*}{}
\author[affiliation={1,3}]{Mohd Mujtaba}{Akhtar*} 
\author[affiliation={4}]{Swarup Ranjan}{Behera}
\author[affiliation={4}]{Priyabrata}{Mallick}
\author[affiliation={5}]{Pailla Balakrishna}{Reddy}
\author[affiliation={1}]{Arun Balaji}{Buduru}
\author[affiliation={6,7}]{Rajesh}{Sharma}

\affiliation{}{IIIT-Delhi}{India}
\affiliation{}{UPES}{India}
\affiliation{}{V.B.S.P.U}{India}
\affiliation{}{Independent Researcher}{India}
\affiliation{}{Reliance AI}{India}
\affiliation{}{University of Tartu}{Estonia}
\affiliation{}{Plaksha University}{India}
\email{\textcolor{blue}{\texttt{Correspondence:}} orchidp@iiitd.ac.in} 

\usepackage{comment}

\begin{document}

\maketitle
\begingroup
  \renewcommand{\thefootnote}{\fnsymbol{footnote}}
  \setcounter{footnote}{0}
   \footnotetext{* Contributed equally as a first authors.}
\endgroup

\begin{abstract}
\noindent In this work, we introduce the task of singing voice deepfake source attribution (SVDSA). We hypothesize that multimodal foundation models (MMFMs) such as ImageBind, LanguageBind will be most effective for SVDSA as they are better equipped for capturing subtle source-specific characteristics—such as unique timbre, pitch manipulation, or synthesis artifacts of each singing voice deepfake source due to their cross-modality pre-training. Our experiments with MMFMs, speech foundation models and music foundation models verify the hypothesis that MMFMs are the most effective for SVDSA. Furthermore, inspired from related research, we also explore fusion of foundation models (FMs) for improved SVDSA. To this end, we propose a novel framework, \texttt{\textbf{COFFE}} which employs Chernoff Distance as novel loss function for effective fusion of FMs. Through \texttt{\textbf{COFFE}} with the symphony of MMFMs, we attain the topmost performance in comparison to all the individual FMs and baseline fusion methods. 
\end{abstract}

\section{Introduction}
\textit{"Imagine discovering a new song by your favorite artist, only to learn they never recorded it."} With generative technologies advancing at an unprecedented pace, this scenario is no longer hypothetical. Singing voice deepfakes (SVDs) have evolved to a level where they can convincingly mimic an artist’s vocal timbre, seamlessly intertwining \textit{speech articulation} with \textit{musical tonality} \cite{zang2024singfake}. While these innovations open new frontiers in creative expression, they also introduce profound challenges related to authenticity, intellectual property rights, and the ethical deployment of AI in music generation. As deepfake synthesis becomes increasingly sophisticated, {the challenge extends beyond mere detection - understanding the provenance of synthetic audio is now imperative}. Tracing the origins of SVDs has become crucial for safeguarding artistic integrity and mitigating the risks of misuse, yet this remains an uncharted problem in singing voice deepfake forensics.\newline 
While Singing Voice Deepfake Detection (SVDD) has seen notable advancements \cite{zang2024singfake, zang24_interspeech, zhang2024svdd}, the equally critical challenge of source attribution - identifying which model or method generated a deepfake - remains largely unexplored. In this work, we introduce the task of singing voice deepfake source attribution (SVDSA). Unlike conventional deepfake detection, which merely classifies an audio clip as real or fake, source attribution seeks to trace its origin, revealing the generative process behind its creation \cite{SA3, SA1}. Source attribution of deepfake speech has captured significant attention in the research community in recent years \cite{zhang2024distinguishing} in contrast to SVDSA. Muller et al. \cite{muller22b_interspeech} used an RNN-based approach for characterizing seen and unseen speech deepfake source signatures. Further, Klein et al. \cite{klein24_interspeech} and \cite{bhagtani2024attribution} has shown the potential of using state-of-the-art (SOTA) speech foundation models (SFMs) such as wav2vec2, Whisper for speech deepfake source attribution (SDSA). These foundation models (FMs) provides sufficient performance benefit as well as take away the need of training models from scratch. As such these FMs have not only captured attention for SDSA but also for speech deepfake detection \cite{chetia-phukan-etal-2024-heterogeneity}, SVDD \cite{chen24o_interspeech} and so on. \newline
So, as the primary research in SVDSA, we explore various FMs and \textit{we hypothesize that multimodal foundation models (MMFMs), such as ImageBind (IB) and LanguageBind (LB) will be most effective for SVDSA as they are particularly well-suited for capturing subtle, source-specific characteristics—such as unique timbre, pitch variations, and synthesis artifacts—present in singing voice deepfake sources. This advantage arises from their cross-modality pretraining, which enables them to learn rich, complementary representations by leveraging diverse contextual and acoustic information.} To validate our hypothesis, we perform a large-scale comparison of MMFMs, SFMs, and music foundation models (MFMs). We consider SFMs and MFMs in our study as research on SVDD has shown their efficacy \cite{zang2024singfake, zhang2024svdd}. Further, motivated by previous research in various related tasks such as speech deepfake detection \cite{chetia-phukan-etal-2024-heterogeneity}, speech recognition \cite{arunkumar22b_interspeech} as well as SVDD \cite{chen24o_interspeech, guragain2024speech} where fusion of FMs have shown improved performance due to the emergence of complementary behavior of the FMs, we also explore the fusion of FMs for SVDSA. To our end, we propose a novel framework, \texttt{\textbf{COFFE}} (Fusion using \texttt{\textbf{C}}hern\texttt{\textbf{OFF}} Distanc\texttt{\textbf{E})}, for effective fusion of FMs. It leverages Chernoff Distance as a novel loss function for aligning the FMs to a joint feature space. With \texttt{\textbf{COFFE}} through the fusion of LB and IB, we obtain the topmost performance in comparison to all the individual FMs, baseline fusion techniques and setting SOTA in benchmark SVD dataset for future research in SVDSA. \par

\noindent \textbf{To summarize, the main contributions are as follows:}  
\begin{itemize}
    \item We introduce SVDSA, pioneering the task of tracing the generative origins of synthetic singing voices.  
    \item We demonstrate the effectiveness of MMFMs, which outperform unimodal SFMs and MFMs for SVDSA due to their multimodal pre-training.
    \item We propose a novel framework, \texttt{\textbf{COFFE}} for fusion of FMs which uses Chernoff Distance (CD) as novel loss function.  
    \item Using \texttt{\textbf{COFFE}} with fusion of LB and IB we achieved the topmost performance in comparison to individual FMs and baseline fusion techniques. 
    \item We establish the first benchmark for SVDSA.
\end{itemize}
\noindent The code and models proposed in this study are available at: \url{https://github.com/Helix-IIIT-Delhi/COFFE-Singing_Voice_Deepfake}


\begin{figure}[!bt]
    \centering
    \includegraphics[width=0.7\linewidth]{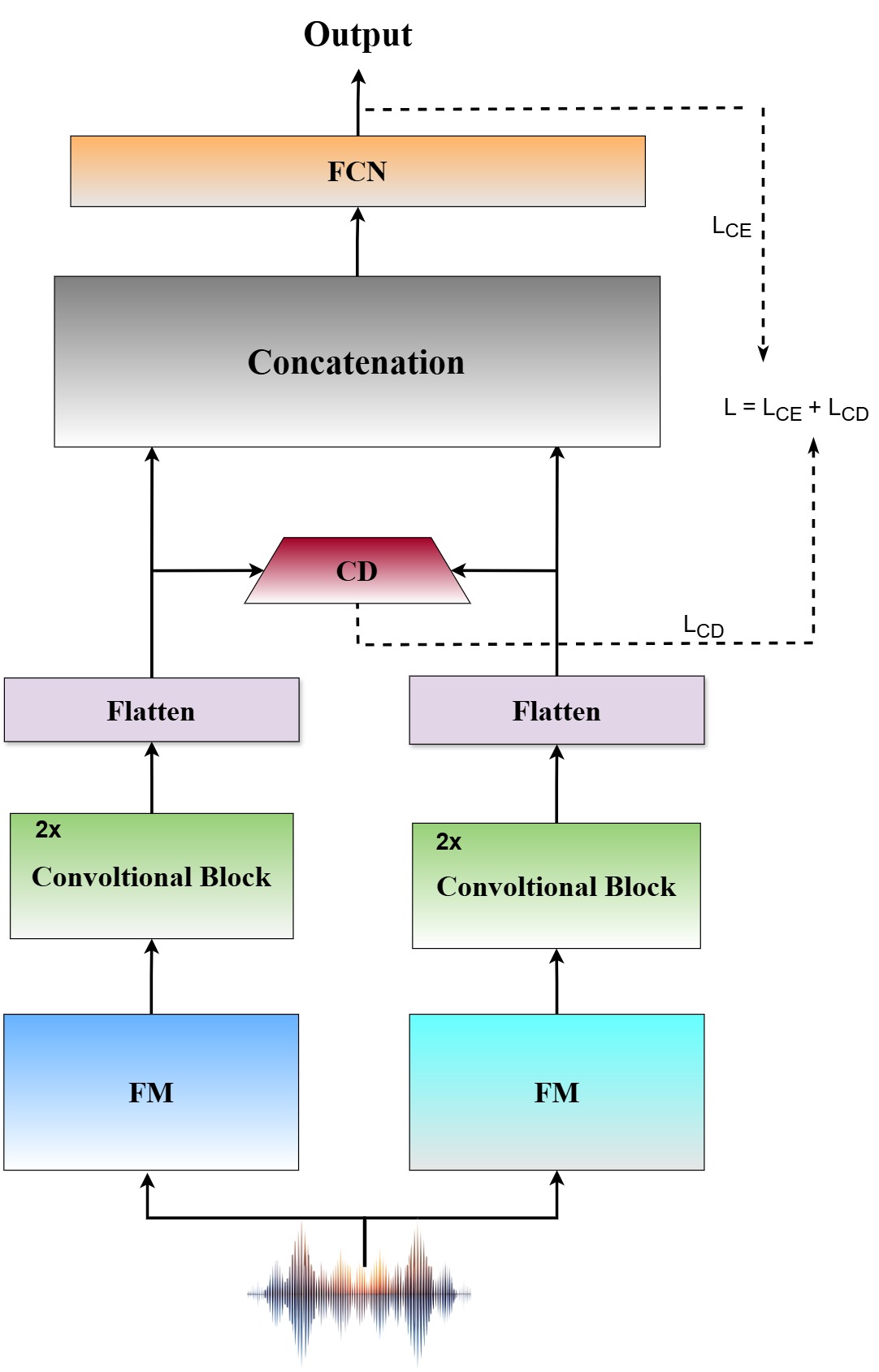}
    \caption{Proposed Modeling Framework for fusion of FMs: \texttt{\textbf{COFEE}}; CD represents the Chernoff Distance; $L_{CD}$, $L_{CE}$, $L$ stands for Chernoff Distance Loss, Cross-entropy loss, Total loss respectively}
    \label{architecture}
\end{figure}

\section{Foundation Models}
In this section, we first discuss the SFMs followed by MFMs and MMFMs considered in our study.

\noindent\textbf{Speech Foundation Models}: We consider WavLM\footnote{\url{https://huggingface.co/facebook/wav2vec2-base}} \cite{baevski2020wav2vec} and Unispeech-SAT\footnote{\url{https://huggingface.co/microsoft/unispeech-sat-base}} \cite{chen2022unispeech} which are SOTA SFMs in SUPERB. Unispeech-SAT incorporates contrastive utterance-wise loss and speaker-aware training while WavLM does its pre-training through masked speech modeling and denoising. We consider the base versions of both WavLM and Unispeech-SAT of 94.70M and 94.68M parameters and pre-trained on 960 hours of english librispeech. We also consider Wav2vec2\footnote{\url{https://huggingface.co/facebook/wav2vec2-base}} \cite{baevski2020wav2vec}. Wav2vec2 is a SFM that is trained in self-supervised manner and applies contrastive learning to masked speech inputs. We use the base version of Wav2vec2 with 95.04M parameters with pre-training done on english librispeech 960 hours. Further, we included SOTA multilingual SFMs such as XLS-R\footnote{\url{https://huggingface.co/facebook/wav2vec2-xls-r-300m}} \cite{babu22_interspeech}, Whisper\footnote{\url{https://huggingface.co/openai/whisper-base}} \cite{radford2023robust}, and MMS\footnote{\url{https://huggingface.co/facebook/mms-1b}} \cite{pratap2024scaling} in our study. XLS-R and MMS are built on top of Wav2vec2 architecture, while XLS-R being trained on 128 languages, MMS extends pre-training to 1406 languages. Whisper is vanilla transformer encoder-decoder architecture and pre-trained in a multi-task learning weakly supervised manner. We consider XLS-R, Whisper, MMS of 300M, 74M and 1B parameters variants. Additionally, we included x-vector\footnote{\url{https://huggingface.co/speechbrain/spkrec-xvect-voxceleb}} \cite{8461375}, a time-delay neural network trained for speaker recognition of 4.2M parameters. We consider it as it has shown effective performance in synthetic speech detection \cite{chetia-phukan-etal-2024-heterogeneity} and we thought it might be helpful for SVDSA. \newline
\noindent \textbf{Music Foundation Models}: MERT series \cite{li2023mert} represents a SOTA MFMs, specifically designed for providing intricate features from musical audio. These MFMs demonstrate exceptional performance across various music-related tasks, including instrument classification, singer indentification, emotion score prediction and so on due to their extensive pretraining on diverse music datasets. We incorporate multiple MERT variants, namely \textit{MERT-v1-330M}\footnote{\url{https://huggingface.co/m-a-p/MERT-v1-330M}}, \textit{MERT-v1-95M}\footnote{\url{https://huggingface.co/m-a-p/MERT-v1-95M}}, \textit{MERT-v0-public}\footnote{\url{https://huggingface.co/m-a-p/MERT-v0-public}}, and \textit{MERT-v0}\footnote{\url{https://huggingface.co/m-a-p/MERT-v0}}. Additionally, we include music2vec-v1 \cite{Music2VecAS}\footnote{\url{https://huggingface.co/m-a-p/music2vec-v1}} which follows a self-supervised training approach, allowing it to effectively capture the nuanced characteristics of music and provide generalized representations for diverse music information retrieval tasks. All MFMs excluding MERT-v1-330M are of 95M parameters each and MERT-v1-330M is of 330M parameters.  \newline
\noindent \textbf{Multimodal Foundation Models}: IB\footnote{\url{https://github.com/facebookresearch/ImageBind/tree/main}} \cite{girdhar2023imagebind} is a MMFM that aligns diverse inputs such as images, audio, text, IMU, depth, and thermal data to a shared image representation space during its pre-training stage by employing an InfoNCE-based optimization strategy. Without requiring explicitly paired training data, IB shows strong generalization across modalities. Similarly, LanguageBind\footnote{\url{https://github.com/PKU-YuanGroup/LanguageBind}} \cite{zhu2023languagebind} aligns multiple data types—including video, depth, audio, and infrared—by anchoring them to a fixed language encoder through contrastive learning. \newline
All FMs requires the input audio to be resampled at different rates: MERT-v1-330M and MERT-v1-95M operate at 24 kHz, whereas MERT-v0-public, MERT-v0, and music2vec-v1 process audio at 16 kHz. Both SFMs and MMFMs necessitate resampling audio to 16 kHz. For feature extraction, average pooling is applied to the final hidden layer of each frozen FM. music2vec-v1 and MERT variants yield 768-dimensional representations, except for MERT-v1-330M, which generates 1024-dimensional representations. Additionally, x-vector, Whisper produces 512-dimensional representations, while IB and LB output 1024 and 768 dimensions, respectively. For whisper, only the encoder is used. XLS-R, MMS produces 1280-dimension representations while Unispeech-SAT, WavLM, Wav2vec2 generates 768-dimensional representations.

\section{Modeling}
In this section, we first discuss the downstream network with individual FMs followed by our proposed framework for fusion of FMs, \texttt{\textbf{COFFE}}. We implemented two distinct downstream for individual FMs—Fully Connected Network (FCN) and CNN. The CNN consists of two 1D convolutional layers with 64 and 128 filters (kernel size = 3), each followed by max pooling (pool size = 2). The features are then flattened and passed through a FCN block with a dense layer of 128. The output layer comprised 8 neurons with softmax as activation function that provides probabilities of SVD sources. The FCN model consisted the same modeling details as the FCN block in CNN.

\subsection{COFFE}
We propose a novel framework, \textbf{\texttt{COFFE}} for the fusion of FMs. The modeling architecture is given in Figure \ref{architecture}. First, the representations from the FMs are passed through two convolutional blocks consisting of 1D convolutional layers with maxpooling with the same modeling as done with individual FMs above. The features are then flattened. We use chernoff distance as novel loss function for aligning the representation space of two FMs. CD is beneficial as it effectively minimizes the separability between feature distributions of the FMs. The CD between two FMs feature space $p$ and $q$, defined as: 

\begin{equation}
L_{CD} = -\log \left( \sum_{i} p_i^s \cdot q_i^{1-s} \right)
\end{equation}

\noindent where $s$ balances their contributions. A higher CD value indicates greater separability and we aim for optimizing it to minimum for aligning the FMs. The features are then concatenated and finally passed through a FCN block with 128 neurons followed by a output layer with 8 neurons as the singfake source classes. Finally, we obtain the total loss function $L$ that combines the $L_{CD}$ and $L_{CE}$ for joint optimization. $L$ is given as: $\mathcal{L} = \mathcal{L}_{CE} + \lambda \cdot \mathcal{L}_{CD} $, where $L_{CE}$, $L_{CD}$ are cross-entropy and CD loss. $\lambda$ is a hyperparameter. The number of trainable parameters varies between 3M and 8M, depending on the size of the FMs representations.
\section{Experiment}

\subsection{Dataset}
We utilized the CtrSVDD \cite{svddchallenge2024}, a benchmark dataset specifically designed for SVDD and the audio samples are in Chinese and Japanese. 
We only consider the synthetic samples for our experiments and it includes 188,486 clips, totaling 260.34 hours. The synthetic samples are generated through 14 distinct synthesis methods (A01–A14). The dataset contains its own official split into train, dev and eval. As train and dev contains the samples generated by the same systems ranging from A01-A08, we used them as training and testing for our source classification models. \par

\noindent\textbf{Training and Hyperparameter Details}: We keep the training epochs as 50 and use Adam as optimizer with cross-entropy as loss function. We keep the learning rate as 1e-3 and use dropout and early dropping for preventing overfitting. For the experiments with \textbf{\texttt{COFEE}}, we keep the value of s and $\lambda$ as 0.3 and 0.1 after some initial experimentation. 

\begin{table}[hbt!]
\centering
\setlength{\tabcolsep}{6pt} 
\tiny
\begin{adjustbox}{max width=\textwidth}
\scriptsize
\centering
\begin{tabular}{l|ccc|ccc}
\toprule
\multirow{2}{*}{\textbf{FM}} & \multicolumn{3}{c|}{\textbf{FCN}}        & \multicolumn{3}{c}{\textbf{CNN}}        \\ \cmidrule(lr){2-4} \cmidrule(lr){5-7}
                       & \textbf{Acc} \(\uparrow\)     & \textbf{F1} \(\uparrow\)      & \textbf{EER} \(\downarrow\)    & \textbf{Acc} \(\uparrow\)     & \textbf{F1} \(\uparrow\)      & \textbf{EER} \(\downarrow\)     \\ \midrule
UNI                    & \cellcolor{emerald5}45.56   & \cellcolor{emerald5}42.22   & \cellcolor{emerald5}22.25   & \cellcolor{emerald5}48.76   & \cellcolor{emerald5}46.56   & \cellcolor{emerald5}18.95   \\
W2V2                   & \cellcolor{emerald10}59.17   & \cellcolor{emerald10}52.12   & \cellcolor{emerald5}17.12   & \cellcolor{emerald10}64.03   & \cellcolor{emerald10}57.90   & \cellcolor{emerald10}13.66   \\
WM                     & \cellcolor{emerald5}35.22   & \cellcolor{emerald5}31.26   & \cellcolor{emerald5}27.87   & \cellcolor{emerald5}38.96   & \cellcolor{emerald5}36.91   & \cellcolor{emerald5}20.18   \\
XL                     & \cellcolor{emerald15}74.90   & \cellcolor{emerald10}64.03   & \cellcolor{emerald10}10.85   & \cellcolor{emerald15}77.51   & \cellcolor{emerald15}69.73   & \cellcolor{emerald10}8.73    \\
WS                     & \cellcolor{emerald10}67.92   & \cellcolor{emerald10}64.69   & \cellcolor{emerald10}12.32   & \cellcolor{emerald10}72.65   & \cellcolor{emerald10}65.11   & \cellcolor{emerald10}10.70   \\
MMS                    & \cellcolor{emerald15}76.36   & \cellcolor{emerald15}67.56   & \cellcolor{emerald15}8.97    & \cellcolor{emerald15}80.41   & \cellcolor{emerald15}76.83   & \cellcolor{emerald15}7.31    \\
XC                     & \cellcolor{emerald10}63.48   & \cellcolor{emerald10}61.17   & \cellcolor{emerald10}13.04   & \cellcolor{emerald10}66.52   & \cellcolor{emerald10}63.95   & \cellcolor{emerald10}12.32   \\
\textbf{LB}          & \cellcolor{emerald15}\textbf{79.69} & \cellcolor{emerald15}\textbf{77.26} & \cellcolor{emerald15}\textbf{6.98}  & \cellcolor{emerald15}\textbf{82.37} & \cellcolor{emerald15}\textbf{79.80} & \cellcolor{emerald15}\textbf{5.35}  \\
IB                     & \cellcolor{emerald15}78.50   & \cellcolor{emerald15}74.41   & \cellcolor{emerald15}7.54   & \cellcolor{emerald15}81.92   & \cellcolor{emerald15}77.90   & \cellcolor{emerald15}6.19   \\
Mv1                    & \cellcolor{emerald5}40.87   & \cellcolor{emerald5}37.02   & \cellcolor{emerald5}25.72   & \cellcolor{emerald5}47.13   & \cellcolor{emerald5}42.98   & \cellcolor{emerald5}19.79   \\
M95                    & \cellcolor{emerald5}47.95   & \cellcolor{emerald5}45.12   & \cellcolor{emerald5}21.44   & \cellcolor{emerald5}50.48   & \cellcolor{emerald5}47.85   & \cellcolor{emerald5}18.71   \\
Mpub                   & \cellcolor{emerald5}53.03   & \cellcolor{emerald5}47.19   & \cellcolor{emerald5}19.52   & \cellcolor{emerald5}55.66   & \cellcolor{emerald5}52.49   & \cellcolor{emerald5}17.77   \\
M330                   & \cellcolor{emerald10}58.64   & \cellcolor{emerald10}54.93   & \cellcolor{emerald10}15.95   & \cellcolor{emerald10}67.96   & \cellcolor{emerald10}63.88   & \cellcolor{emerald10}12.19   \\
Mv0                    & \cellcolor{emerald5}45.47   & \cellcolor{emerald5}42.81   & \cellcolor{emerald5}24.58   & \cellcolor{emerald5}48.55   & \cellcolor{emerald5}45.53   & \cellcolor{emerald5}19.65   \\
\bottomrule
\end{tabular}
\end{adjustbox}
\caption{Evaluation Scores for different FMs; Abbreviations used: \textbf{UNI}: Unispeech-SAT, \textbf{W2V2}: Wav2vec2, \textbf{WM}: WavLM, \textbf{WS}: Whisper, \textbf{XL}: XLS-R, \textbf{XC}: x-vector, \textbf{MMS}: MMS, \textbf{HUB}: HuBERT, \textbf{LB}: LanguageBind, \textbf{IB}: ImageBind, \textbf{Mv1}: music2vec-v1, \textbf{M95}: MERT-v1-95M, \textbf{Mpub}: MERT-v0-public, \textbf{M330}: MERT-v1-330M, \textbf{Mv0}: MERT-v0; Scores are expressed as percentages (\%); The abbreviations and information in this Table \ref{tab:model_performancesingle} are kept same for Table \ref{tab:complete_performance}}
\label{tab:model_performancesingle}
\end{table}

\begin{figure}[]
    \centering
    \subfloat[]{%
        \includegraphics[width=0.23\textwidth]{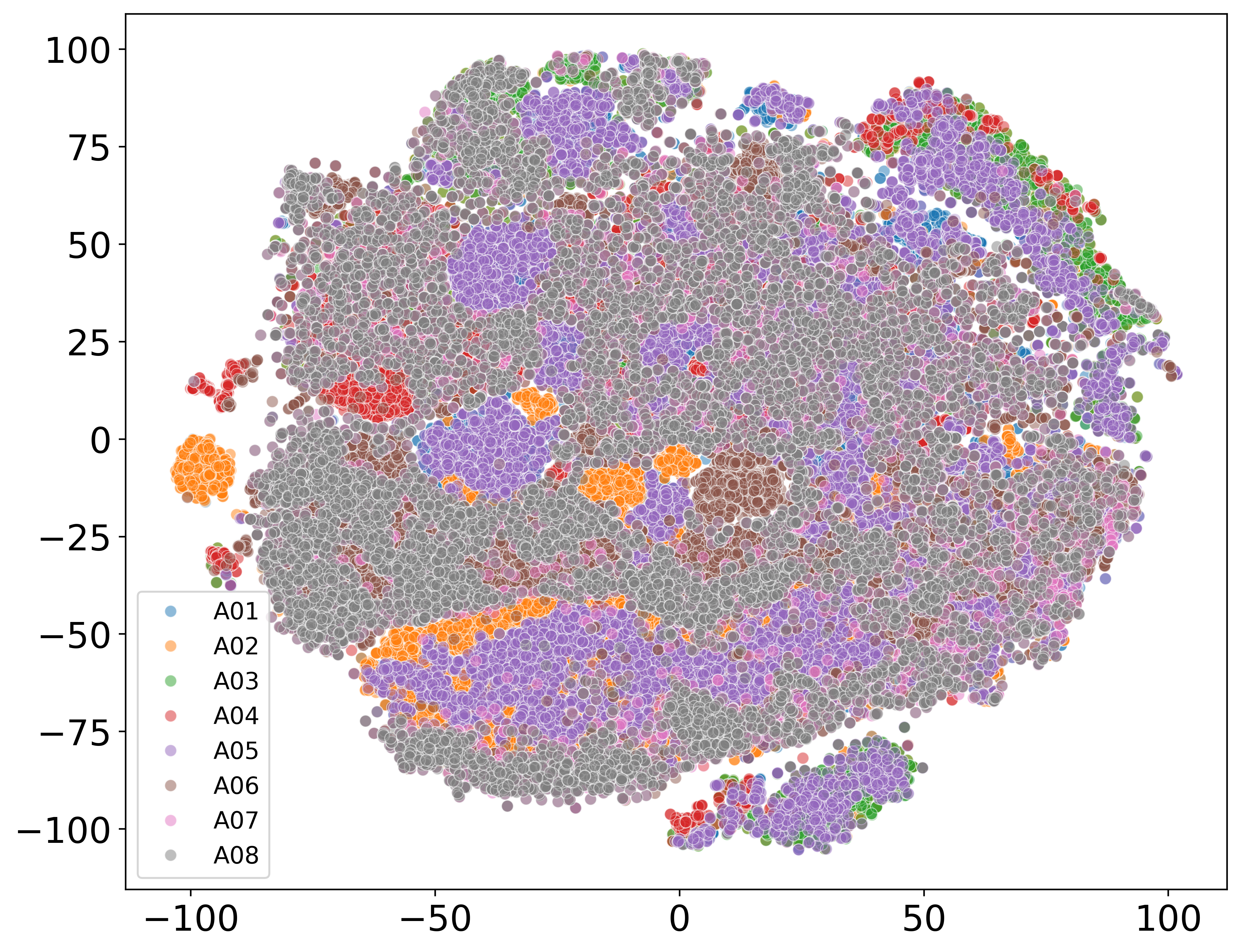}
    }
    \hfill
    \subfloat[]{%
        \includegraphics[width=0.23\textwidth]{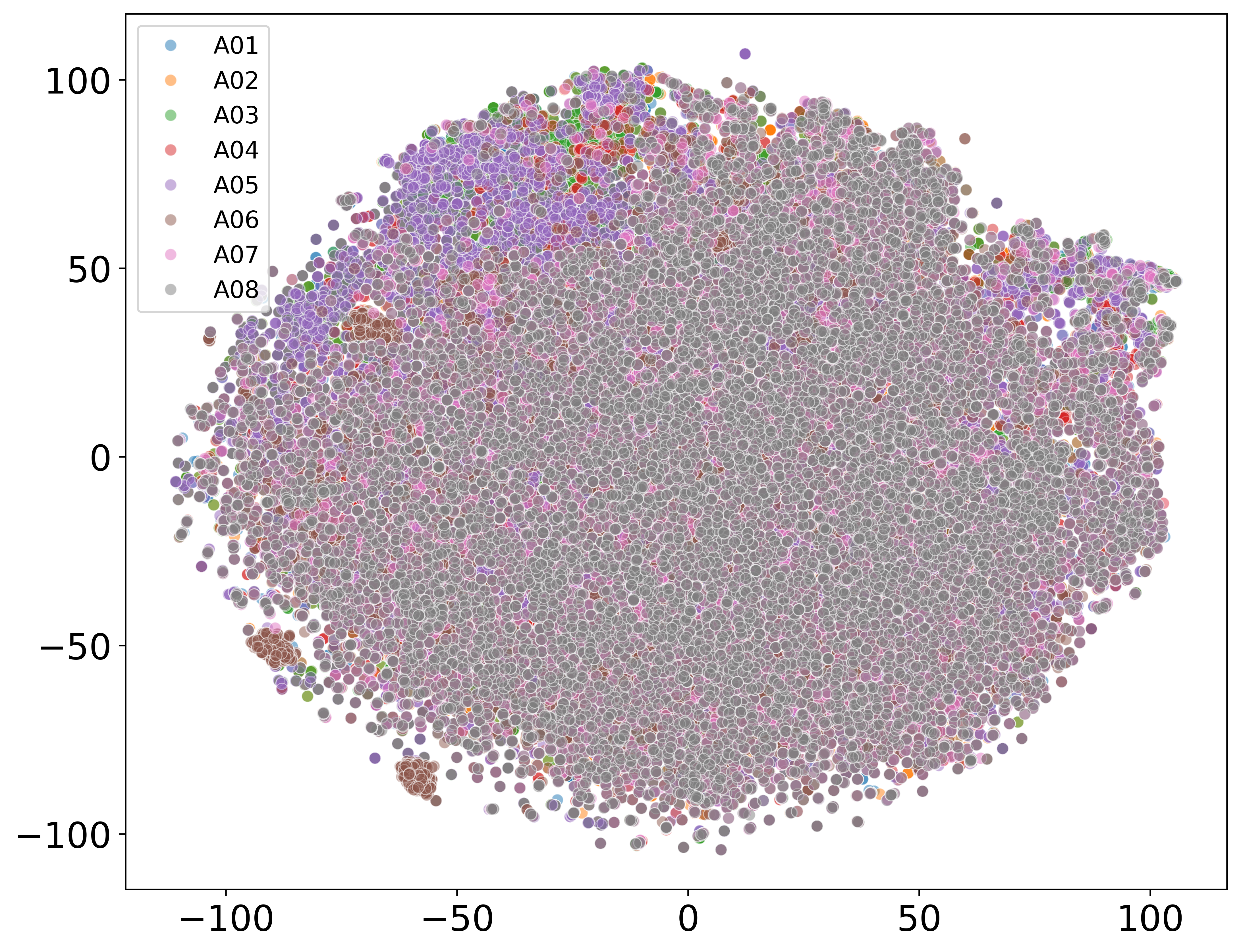}
    }
    \hfill
    \subfloat[]{%
        \includegraphics[width=0.23\textwidth]{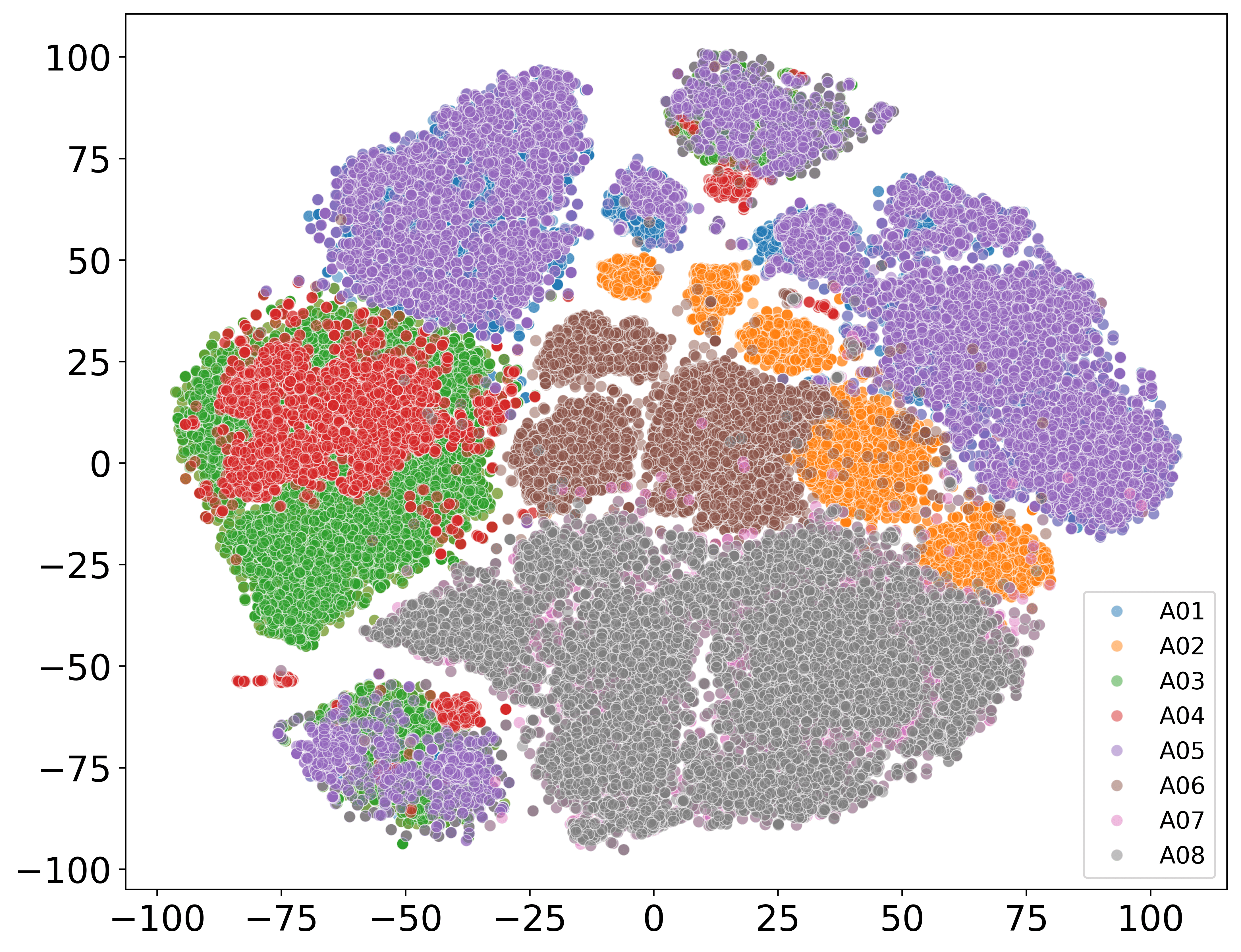}
    }
    \hfill
    \subfloat[]{%
        \includegraphics[width=0.23\textwidth]{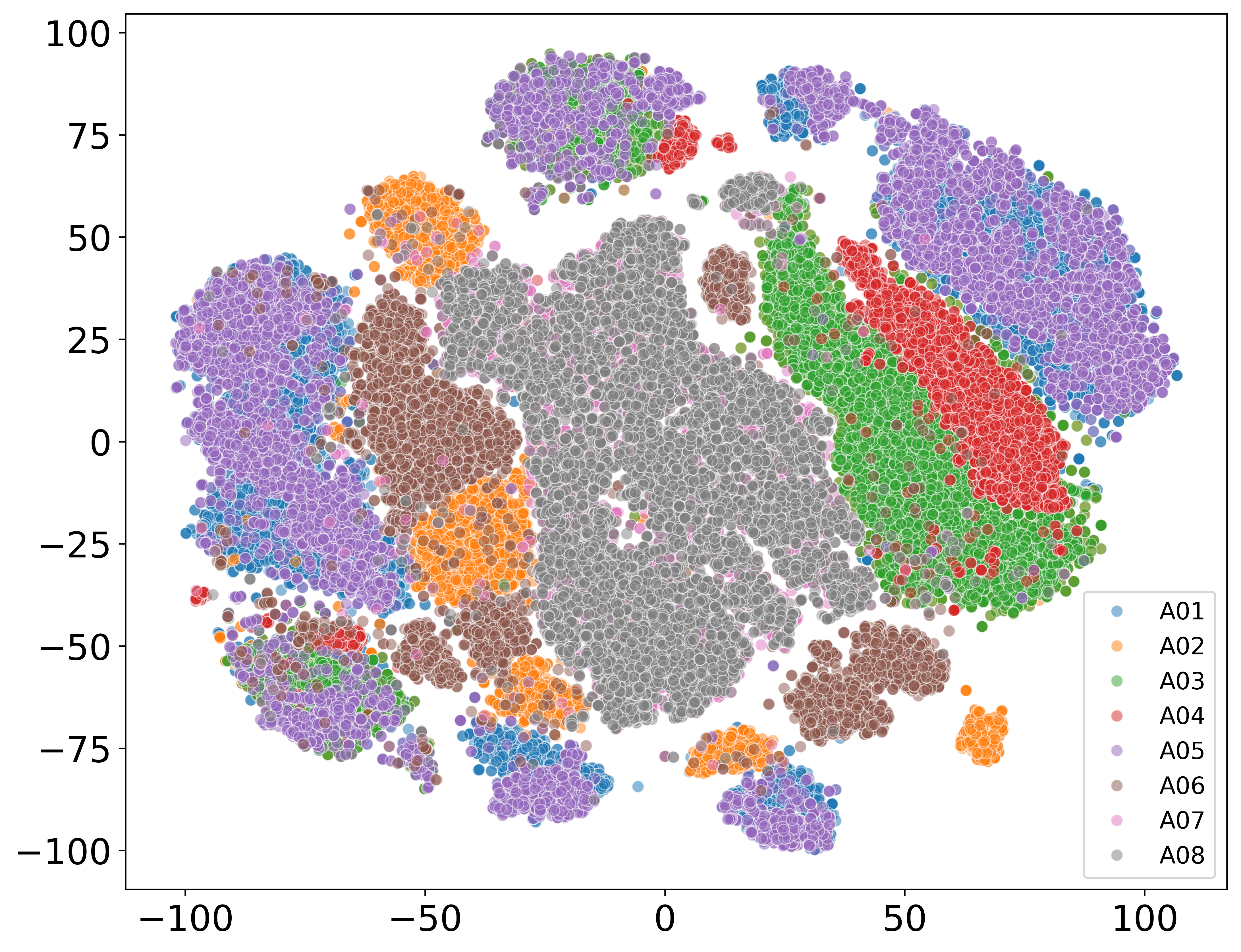}
    }
    \caption{t-SNE Plots- (a) XLS-R (b) MERT-v1-330M (c) LanguageBind (d) ImageBind}
    \label{fig:tsne}
\end{figure}

\begin{figure}[!bt]
    \centering
    \subfloat[]{%
        \includegraphics[width=0.2\textwidth]{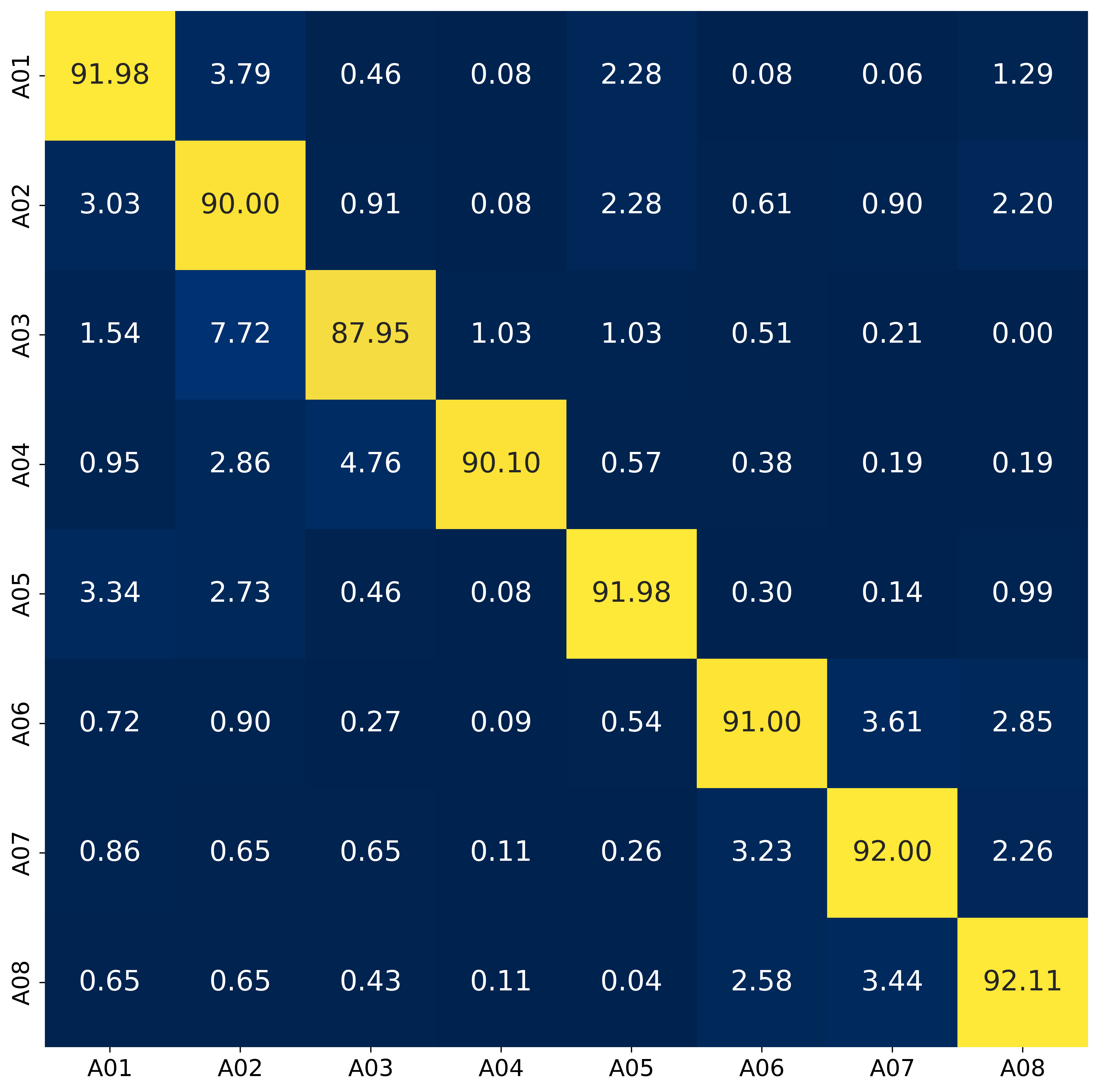}
    }
    \hfill
    \subfloat[]{%
        \includegraphics[width=0.2\textwidth]{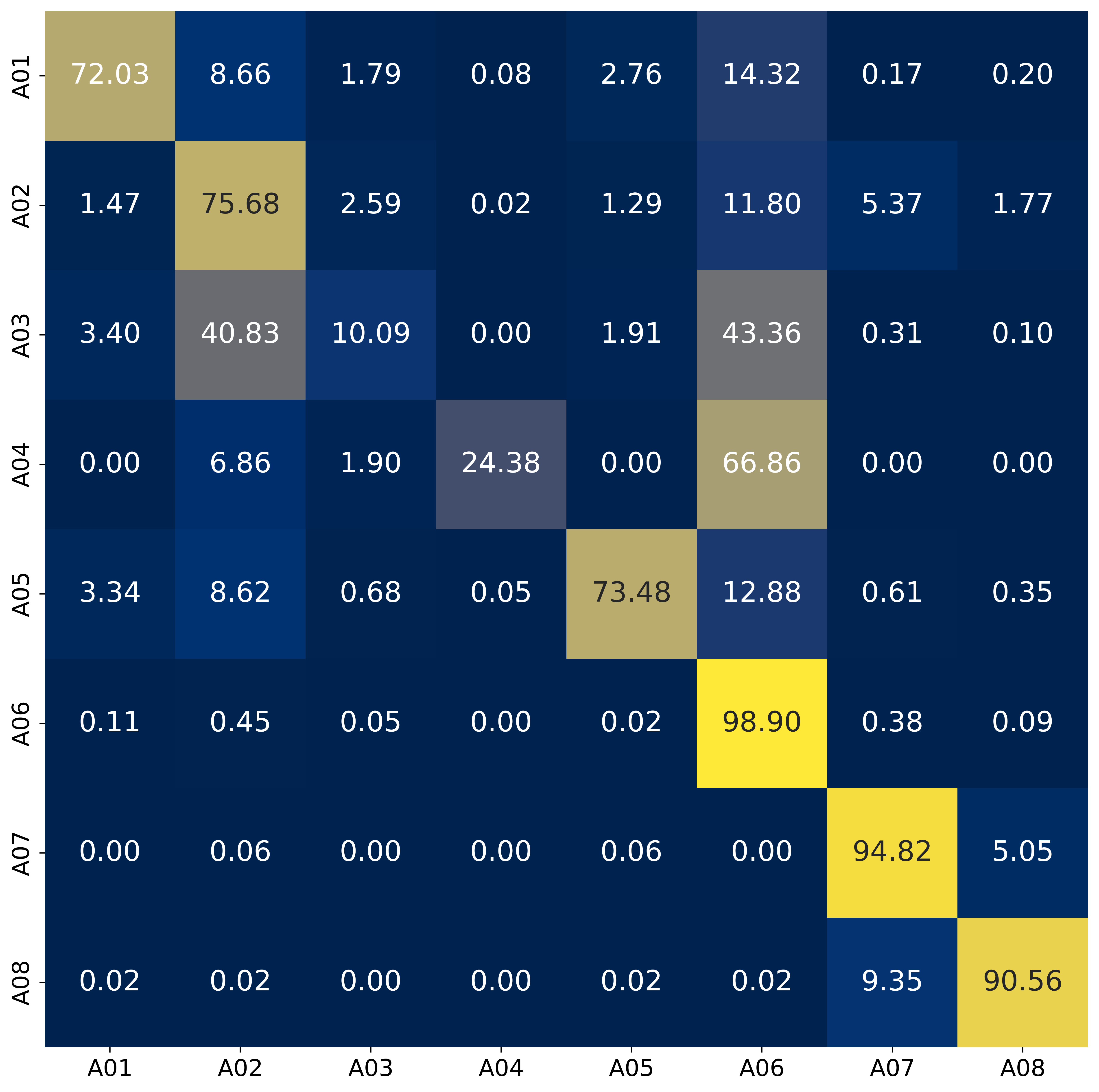}
    }
    \caption{Confusion matrices: (a) \textbf{\texttt{COFEE(LB + IB)}} (b) CNN(LB); The x-axis represents the predicted values, while the y-axis denotes the true values}
    \label{fig:CM}
\end{figure}

\begin{table}[]
\setlength{\tabcolsep}{4pt}
\scriptsize
\centering
\begin{adjustbox}{max width=\textwidth}
\scriptsize
\centering
\begin{tabular}{l|ccc|ccc}
\toprule
& \multicolumn{3}{c|}{\textbf{Concatenation}} & \multicolumn{3}{c}{\textbf{\texttt{COFFE}}} \\
\cmidrule(lr){2-4} \cmidrule(lr){5-7}
\textbf{Combinations} & \textbf{ACC} \(\uparrow\) & \textbf{F1} \(\uparrow\) & \textbf{EER} \(\downarrow\) & \textbf{ACC} \(\uparrow\) & \textbf{F1} \(\uparrow\) & \textbf{EER} \(\downarrow\) \\
\midrule
UNI + W2V2     & \cellcolor{emerald10}61.78 & \cellcolor{emerald10}54.02 & \cellcolor{emerald5}17.13 & \cellcolor{emerald10}62.36 & \cellcolor{emerald10}61.96 & \cellcolor{emerald5}15.20 \\
UNI + WM       & \cellcolor{emerald5}52.33  & \cellcolor{emerald5}44.28  & \cellcolor{emerald5}23.11 & \cellcolor{emerald5}54.96  & \cellcolor{emerald5}53.61  & \cellcolor{emerald5}19.63 \\
UNI + XL       & \cellcolor{emerald15}74.60 & \cellcolor{emerald10}61.24 & \cellcolor{emerald10}11.97 & \cellcolor{emerald10}72.31 & \cellcolor{emerald15}78.50 & \cellcolor{emerald10}7.39  \\
UNI + WS       & \cellcolor{emerald10}68.49 & \cellcolor{emerald10}62.67 & \cellcolor{emerald10}13.04 & \cellcolor{emerald10}73.18 & \cellcolor{emerald10}71.37 & \cellcolor{emerald10}9.62  \\
UNI + MMS      & \cellcolor{emerald10}73.23 & \cellcolor{emerald15}68.49 & \cellcolor{emerald10}9.16  & \cellcolor{emerald10}73.31 & \cellcolor{emerald15}78.22 & \cellcolor{emerald10}7.94  \\
UNI + XC       & \cellcolor{emerald10}65.89 & \cellcolor{emerald10}58.65 & \cellcolor{emerald10}11.80 & \cellcolor{emerald10}67.34 & \cellcolor{emerald10}66.34 & \cellcolor{emerald10}8.61  \\
UNI + LB       & \cellcolor{emerald15}78.86 & \cellcolor{emerald15}74.00 & \cellcolor{emerald15}8.22  & \cellcolor{emerald15}80.91 & \cellcolor{emerald15}79.36 & \cellcolor{emerald15}5.34  \\
UNI + IB       & \cellcolor{emerald10}72.10 & \cellcolor{emerald10}57.77 & \cellcolor{emerald10}12.65 & \cellcolor{emerald10}76.36 & \cellcolor{emerald10}74.36 & \cellcolor{emerald10}10.37 \\
UNI + Mv1      & \cellcolor{emerald10}59.28 & \cellcolor{emerald10}52.29 & \cellcolor{emerald5}19.09  & \cellcolor{emerald5}47.13  & \cellcolor{emerald10}42.98 & \cellcolor{emerald5}16.43 \\
UNI + M95      & \cellcolor{emerald10}63.64 & \cellcolor{emerald10}57.26 & \cellcolor{emerald5}16.69  & \cellcolor{emerald10}68.36 & \cellcolor{emerald10}67.63 & \cellcolor{emerald10}11.91 \\
UNI + Mpub     & \cellcolor{emerald10}62.83 & \cellcolor{emerald10}55.81 & \cellcolor{emerald10}14.97 & \cellcolor{emerald10}66.95 & \cellcolor{emerald10}65.09 & \cellcolor{emerald10}9.33  \\
UNI + M330     & \cellcolor{emerald10}66.74 & \cellcolor{emerald10}60.54 & \cellcolor{emerald10}13.63 & \cellcolor{emerald10}71.69 & \cellcolor{emerald10}70.34 & \cellcolor{emerald10}11.53 \\
UNI + Mv0      & \cellcolor{emerald10}60.91 & \cellcolor{emerald10}53.86 & \cellcolor{emerald5}18.35  & \cellcolor{emerald10}63.85 & \cellcolor{emerald10}62.31 & \cellcolor{emerald10}13.31 \\
W2V2 + WM      & \cellcolor{emerald10}60.06 & \cellcolor{emerald10}52.18 & \cellcolor{emerald5}16.35  & \cellcolor{emerald10}65.85 & \cellcolor{emerald10}63.29 & \cellcolor{emerald10}11.61 \\
W2V2 + XL      & \cellcolor{emerald10}73.05 & \cellcolor{emerald10}65.89 & \cellcolor{emerald10}10.13  & \cellcolor{emerald10}74.31 & \cellcolor{emerald10}73.35 & \cellcolor{emerald15}6.39  \\
W2V2 + WS      & \cellcolor{emerald10}71.24 & \cellcolor{emerald10}65.52 & \cellcolor{emerald10}11.75  & \cellcolor{emerald10}73.36 & \cellcolor{emerald10}72.39 & \cellcolor{emerald10}9.32  \\
W2V2 + MMS     & \cellcolor{emerald10}71.78 & \cellcolor{emerald10}62.73 & \cellcolor{emerald10}10.58  & \cellcolor{emerald10}74.62 & \cellcolor{emerald10}71.59 & \cellcolor{emerald10}8.62  \\
W2V2 + XC      & \cellcolor{emerald10}68.15 & \cellcolor{emerald10}57.41 & \cellcolor{emerald10}12.65  & \cellcolor{emerald10}75.35 & \cellcolor{emerald10}72.64 & \cellcolor{emerald10}10.37 \\
W2V2 + LB      & \cellcolor{emerald15}77.02 & \cellcolor{emerald15}72.36 & \cellcolor{emerald10}9.54   & \cellcolor{emerald15}81.36 & \cellcolor{emerald15}80.91 & \cellcolor{emerald15}6.32  \\
W2V2 + IB      & \cellcolor{emerald10}71.48 & \cellcolor{emerald10}58.20 & \cellcolor{emerald10}12.14  & \cellcolor{emerald10}76.61 & \cellcolor{emerald10}75.30 & \cellcolor{emerald10}9.33  \\
W2V2 + Mv1     & \cellcolor{emerald10}60.23 & \cellcolor{emerald5}50.59  & \cellcolor{emerald5}17.23  & \cellcolor{emerald5}62.93  & \cellcolor{emerald10}61.39 & \cellcolor{emerald10}13.96 \\
W2V2 + M95     & \cellcolor{emerald10}63.46 & \cellcolor{emerald10}54.13 & \cellcolor{emerald5}16.79  & \cellcolor{emerald10}65.23 & \cellcolor{emerald10}64.94 & \cellcolor{emerald10}10.63 \\
W2V2 + Mpub    & \cellcolor{emerald10}62.51 & \cellcolor{emerald10}52.15 & \cellcolor{emerald5}16.52  & \cellcolor{emerald10}64.36 & \cellcolor{emerald10}63.96 & \cellcolor{emerald10}11.91 \\
W2V2 + M330    & \cellcolor{emerald10}64.98 & \cellcolor{emerald10}55.19 & \cellcolor{emerald10}15.01  & \cellcolor{emerald10}67.96 & \cellcolor{emerald10}66.11 & \cellcolor{emerald10}9.61  \\
W2V2 + Mv0     & \cellcolor{emerald10}61.38 & \cellcolor{emerald5}51.19  & \cellcolor{emerald5}17.13  & \cellcolor{emerald10}63.31 & \cellcolor{emerald10}62.96 & \cellcolor{emerald10}12.33 \\
WM + XL        & \cellcolor{emerald10}73.00 & \cellcolor{emerald10}59.27 & \cellcolor{emerald10}11.06  & \cellcolor{emerald10}75.66 & \cellcolor{emerald10}74.31 & \cellcolor{emerald15}6.97  \\
WM + WS        & \cellcolor{emerald10}67.81 & \cellcolor{emerald10}60.31 & \cellcolor{emerald10}12.87  & \cellcolor{emerald10}69.34 & \cellcolor{emerald10}68.36 & \cellcolor{emerald10}8.63  \\
WM + MMS       & \cellcolor{emerald15}75.87 & \cellcolor{emerald10}66.99 & \cellcolor{emerald10}10.34  & \cellcolor{emerald10}76.93 & \cellcolor{emerald15}77.23 & \cellcolor{emerald10}10.01 \\
WM + XC        & \cellcolor{emerald10}65.41 & \cellcolor{emerald10}58.69 & \cellcolor{emerald10}12.16  & \cellcolor{emerald10}66.31 & \cellcolor{emerald10}64.96 & \cellcolor{emerald10}8.19  \\
WM + LB        & \cellcolor{emerald15}78.48 & \cellcolor{emerald15}73.50 & \cellcolor{emerald10}8.58   & \cellcolor{emerald15}81.63 & \cellcolor{emerald15}80.14 & \cellcolor{emerald10}8.13  \\
WM + IB        & \cellcolor{emerald10}69.41 & \cellcolor{emerald10}59.61 & \cellcolor{emerald10}9.98   & \cellcolor{emerald15}78.54 & \cellcolor{emerald10}70.13 & \cellcolor{emerald10}8.64  \\
WM + Mv1       & \cellcolor{emerald5}54.19  & \cellcolor{emerald5}47.05  & \cellcolor{emerald5}22.41  & \cellcolor{emerald5}56.96  & \cellcolor{emerald5}55.14  & \cellcolor{emerald5}19.17  \\
WM + M95       & \cellcolor{emerald10}58.18 & \cellcolor{emerald5}50.63  & \cellcolor{emerald5}18.84  & \cellcolor{emerald10}61.08 & \cellcolor{emerald10}60.31 & \cellcolor{emerald5}16.42  \\
WM + Mpub      & \cellcolor{emerald10}58.34 & \cellcolor{emerald5}49.78  & \cellcolor{emerald5}16.92  & \cellcolor{emerald10}60.27 & \cellcolor{emerald5}52.67 & \cellcolor{emerald5}16.17  \\
WM + M330      & \cellcolor{emerald10}63.42 & \cellcolor{emerald10}57.45 & \cellcolor{emerald10}14.54  & \cellcolor{emerald10}65.63 & \cellcolor{emerald10}64.49 & \cellcolor{emerald10}11.16  \\
WM + Mv0       & \cellcolor{emerald5}56.46  & \cellcolor{emerald5}49.49  & \cellcolor{emerald5}20.31  & \cellcolor{emerald10}59.61 & \cellcolor{emerald10}58.31 & \cellcolor{emerald5}16.37  \\
XL + WS        & \cellcolor{emerald15}79.41 & \cellcolor{emerald10}68.59 & \cellcolor{emerald10}7.80  & \cellcolor{emerald15}82.64 & \cellcolor{emerald15}81.37 & \cellcolor{emerald15}6.09  \\
XL + MMS       & \cellcolor{emerald15}76.56 & \cellcolor{emerald10}67.21 & \cellcolor{emerald10}8.42  & \cellcolor{emerald15}78.34 & \cellcolor{emerald15}77.31 & \cellcolor{emerald10}8.09  \\
XL + XC        & \cellcolor{emerald15}78.56 & \cellcolor{emerald15}73.96 & \cellcolor{emerald10}10.36 & \cellcolor{emerald15}81.61 & \cellcolor{emerald15}80.67 & \cellcolor{emerald10}8.63  \\
XL + LB        & \cellcolor{emerald15}77.80 & \cellcolor{emerald15}72.40 & \cellcolor{emerald10}8.08  & \cellcolor{emerald15}79.38 & \cellcolor{emerald15}77.31 & \cellcolor{emerald10}8.31  \\
XL + IB        & \cellcolor{emerald15}78.94 & \cellcolor{emerald15}73.50 & \cellcolor{emerald10}10.06 & \cellcolor{emerald15}80.37 & \cellcolor{emerald15}79.09 & \cellcolor{emerald10}10.03 \\
XL + Mv1       & \cellcolor{emerald15}74.07 & \cellcolor{emerald10}66.53 & \cellcolor{emerald10}11.16 & \cellcolor{emerald10}74.93 & \cellcolor{emerald10}73.52 & \cellcolor{emerald10}8.34  \\
XL + M95       & \cellcolor{emerald10}75.04 & \cellcolor{emerald10}60.03 & \cellcolor{emerald10}10.58 & \cellcolor{emerald15}77.31 & \cellcolor{emerald15}75.99 & \cellcolor{emerald10}9.06  \\
XL + Mpub      & \cellcolor{emerald15}76.11 & \cellcolor{emerald10}66.09 & \cellcolor{emerald10}9.61  & \cellcolor{emerald15}78.05 & \cellcolor{emerald10}74.70 & \cellcolor{emerald10}8.20  \\
XL + M330      & \cellcolor{emerald15}75.36 & \cellcolor{emerald10}65.91 & \cellcolor{emerald10}12.29 & \cellcolor{emerald10}76.89 & \cellcolor{emerald10}75.47 & \cellcolor{emerald10}11.37 \\
XL + Mv0       & \cellcolor{emerald10}71.36 & \cellcolor{emerald10}61.11 & \cellcolor{emerald10}11.04 & \cellcolor{emerald10}72.89 & \cellcolor{emerald10}71.67 & \cellcolor{emerald10}9.53  \\
WS + MMS       & \cellcolor{emerald15}77.62 & \cellcolor{emerald10}70.21 & \cellcolor{emerald10}8.24  & \cellcolor{emerald15}81.64 & \cellcolor{emerald15}80.34 & \cellcolor{emerald15}6.32  \\
WS + XC        & \cellcolor{emerald10}73.56 & \cellcolor{emerald15}71.69 & \cellcolor{emerald10}10.96 & \cellcolor{emerald10}75.28 & \cellcolor{emerald10}74.31 & \cellcolor{emerald10}8.94  \\
WS + LB        & \cellcolor{emerald15}78.64 & \cellcolor{emerald15}74.07 & \cellcolor{emerald10}8.36  & \cellcolor{emerald15}81.39 & \cellcolor{emerald15}80.31 & \cellcolor{emerald15}6.17  \\
WS + IB        & \cellcolor{emerald10}72.38 & \cellcolor{emerald10}68.88 & \cellcolor{emerald10}10.87 & \cellcolor{emerald10}76.64 & \cellcolor{emerald10}75.13 & \cellcolor{emerald10}9.84  \\
WS + Mv1       & \cellcolor{emerald10}69.78 & \cellcolor{emerald10}64.32 & \cellcolor{emerald10}12.75 & \cellcolor{emerald10}71.18 & \cellcolor{emerald10}70.61 & \cellcolor{emerald10}9.49  \\
WS + M95       & \cellcolor{emerald10}69.27 & \cellcolor{emerald10}64.83 & \cellcolor{emerald10}11.57 & \cellcolor{emerald10}72.58 & \cellcolor{emerald10}71.23 & \cellcolor{emerald10}9.82  \\
WS + Mpub      & \cellcolor{emerald10}70.48 & \cellcolor{emerald10}65.90 & \cellcolor{emerald10}12.08 & \cellcolor{emerald10}72.71 & \cellcolor{emerald10}69.57 & \cellcolor{emerald10}11.80 \\
WS + M330      & \cellcolor{emerald10}71.77 & \cellcolor{emerald10}67.33 & \cellcolor{emerald10}11.40 & \cellcolor{emerald10}74.56 & \cellcolor{emerald10}73.64 & \cellcolor{emerald10}8.26  \\
WS + Mv0       & \cellcolor{emerald10}69.71 & \cellcolor{emerald10}64.42 & \cellcolor{emerald10}12.72 & \cellcolor{emerald10}73.64 & \cellcolor{emerald10}72.54 & \cellcolor{emerald10}10.67 \\
MMS + XC       & \cellcolor{emerald15}76.54 & \cellcolor{emerald15}71.35 & \cellcolor{emerald10}10.31 & \cellcolor{emerald15}77.96 & \cellcolor{emerald15}76.68 & \cellcolor{emerald10}8.22  \\
MMS + LB       & \cellcolor{emerald15}79.05 & \cellcolor{emerald15}75.43 & \cellcolor{emerald10}9.75  & \cellcolor{emerald15}83.68 & \cellcolor{emerald15}82.27 & \cellcolor{emerald10}9.35  \\
MMS + IB       & \cellcolor{emerald15}76.59 & \cellcolor{emerald15}71.38 & \cellcolor{emerald10}10.06 & \cellcolor{emerald15}78.34 & \cellcolor{emerald15}76.13 & \cellcolor{emerald10}9.08  \\
MMS + Mv1      & \cellcolor{emerald15}74.30 & \cellcolor{emerald15}70.67 & \cellcolor{emerald10}9.87  & \cellcolor{emerald15}76.34 & \cellcolor{emerald15}74.46 & \cellcolor{emerald10}9.03  \\
MMS + M95      & \cellcolor{emerald10}72.98 & \cellcolor{emerald10}63.75 & \cellcolor{emerald10}9.57  & \cellcolor{emerald10}75.38 & \cellcolor{emerald10}74.49 & \cellcolor{emerald10}9.64  \\
MMS + Mpub     & \cellcolor{emerald15}76.13 & \cellcolor{emerald10}67.10 & \cellcolor{emerald15}8.82 & \cellcolor{emerald15}78.93 & \cellcolor{emerald15}77.08 & \cellcolor{emerald15}7.51  \\
MMS + M330     & \cellcolor{emerald15}75.56 & \cellcolor{emerald15}70.77 & \cellcolor{emerald10}8.98  & \cellcolor{emerald15}77.34 & \cellcolor{emerald15}76.31 & \cellcolor{emerald15}7.34  \\
MMS + Mv0      & \cellcolor{emerald15}74.25 & \cellcolor{emerald10}68.40 & \cellcolor{emerald10}11.95 & \cellcolor{emerald15}76.68 & \cellcolor{emerald15}75.39 & \cellcolor{emerald10}10.62 \\
XC + LB        & \cellcolor{emerald15}82.02 & \cellcolor{emerald15}80.15 & \cellcolor{emerald15}6.54  & \cellcolor{emerald15}83.64 & \cellcolor{emerald15}82.66 & \cellcolor{emerald15}4.67  \\
XC + IB        & \cellcolor{emerald10}75.89 & \cellcolor{emerald10}68.45 & \cellcolor{emerald10}11.65 & \cellcolor{emerald15}77.63 & \cellcolor{emerald10}75.58 & \cellcolor{emerald10}9.85  \\
XC + Mv1       & \cellcolor{emerald10}65.23 & \cellcolor{emerald10}63.56 & \cellcolor{emerald5}14.63  & \cellcolor{emerald10}67.73 & \cellcolor{emerald10}66.69 & \cellcolor{emerald10}11.63 \\
XC + M95       & \cellcolor{emerald10}64.23 & \cellcolor{emerald10}57.88 & \cellcolor{emerald10}12.36 & \cellcolor{emerald10}66.38 & \cellcolor{emerald10}65.59 & \cellcolor{emerald10}10.16 \\
XC + Mpub      & \cellcolor{emerald5}61.35  & \cellcolor{emerald10}57.46 & \cellcolor{emerald10}11.36 & \cellcolor{emerald10}63.28 & \cellcolor{emerald10}61.03 & \cellcolor{emerald10}9.34  \\
XC + M330      & \cellcolor{emerald10}64.65 & \cellcolor{emerald10}58.98 & \cellcolor{emerald10}13.65 & \cellcolor{emerald10}66.37 & \cellcolor{emerald10}64.49 & \cellcolor{emerald10}11.44 \\
XC + Mv0       & \cellcolor{emerald5}59.63  & \cellcolor{emerald5}55.41  & \cellcolor{emerald10}11.65 & \cellcolor{emerald10}62.38 & \cellcolor{emerald10}60.34 & \cellcolor{emerald10}10.74 \\
\textbf{LB + IB}        & \cellcolor{emerald15}\textbf{89.62} & \cellcolor{emerald15}\textbf{83.88} & \cellcolor{emerald15}\textbf{3.75}  & \cellcolor{emerald15}\textbf{91.16} & \cellcolor{emerald15}\textbf{90.03} & \cellcolor{emerald15}\textbf{3.63} \\
LB + Mv1       & \cellcolor{emerald15}77.60 & \cellcolor{emerald15}72.30 & \cellcolor{emerald10}8.91  & \cellcolor{emerald15}78.63 & \cellcolor{emerald15}77.62 & \cellcolor{emerald10}8.03  \\
LB + M95       & \cellcolor{emerald15}77.70 & \cellcolor{emerald15}77.20 & \cellcolor{emerald10}8.65  & \cellcolor{emerald15}80.37 & \cellcolor{emerald15}78.86 & \cellcolor{emerald10}7.39  \\
LB + Mpub      & \cellcolor{emerald15}77.63 & \cellcolor{emerald15}72.88 & \cellcolor{emerald10}9.20  & \cellcolor{emerald15}79.78 & \cellcolor{emerald15}76.62 & \cellcolor{emerald10}8.33  \\
LB + M330      & \cellcolor{emerald15}80.40 & \cellcolor{emerald15}77.59 & \cellcolor{emerald15}7.63  & \cellcolor{emerald15}82.29 & \cellcolor{emerald15}81.17 & \cellcolor{emerald15}7.39  \\
LB + Mv0       & \cellcolor{emerald15}79.06 & \cellcolor{emerald15}75.84 & \cellcolor{emerald10}8.04  & \cellcolor{emerald15}82.23 & \cellcolor{emerald15}81.18 & \cellcolor{emerald10}7.93  \\
IB + Mv1       & \cellcolor{emerald10}68.06 & \cellcolor{emerald5}55.21  & \cellcolor{emerald10}13.22 & \cellcolor{emerald10}71.94 & \cellcolor{emerald10}68.83 & \cellcolor{emerald10}11.38 \\
IB + M95       & \cellcolor{emerald10}72.49 & \cellcolor{emerald10}71.56 & \cellcolor{emerald10}14.38 & \cellcolor{emerald10}73.94 & \cellcolor{emerald10}71.28 & \cellcolor{emerald10}11.07 \\
IB + Mpub      & \cellcolor{emerald10}71.47 & \cellcolor{emerald10}57.79 & \cellcolor{emerald5}16.02  & \cellcolor{emerald10}72.29 & \cellcolor{emerald10}71.13 & \cellcolor{emerald10}13.03 \\
IB + M330      & \cellcolor{emerald10}69.95 & \cellcolor{emerald10}57.57 & \cellcolor{emerald10}15.48 & \cellcolor{emerald10}74.24 & \cellcolor{emerald10}71.16 & \cellcolor{emerald10}12.92 \\
IB + Mv0       & \cellcolor{emerald10}68.53 & \cellcolor{emerald10}58.68 & \cellcolor{emerald10}14.89 & \cellcolor{emerald10}70.38 & \cellcolor{emerald10}70.08 & \cellcolor{emerald10}9.36  \\
Mv1 + M95      & \cellcolor{emerald5}57.94  & \cellcolor{emerald5}51.52  & \cellcolor{emerald5}19.77  & \cellcolor{emerald10}61.93 & \cellcolor{emerald10}58.39 & \cellcolor{emerald10}12.83 \\
Mv1 + Mpub     & \cellcolor{emerald5}57.57  & \cellcolor{emerald5}49.67  & \cellcolor{emerald5}18.17  & \cellcolor{emerald10}59.93 & \cellcolor{emerald10}58.81 & \cellcolor{emerald10}13.93 \\
Mv1 + M330     & \cellcolor{emerald10}62.89 & \cellcolor{emerald10}57.46 & \cellcolor{emerald10}13.25  & \cellcolor{emerald10}63.49 & \cellcolor{emerald10}63.02 & \cellcolor{emerald10}11.52 \\
Mv1 + Mv0      & \cellcolor{emerald5}53.54  & \cellcolor{emerald5}46.78  & \cellcolor{emerald5}22.63  & \cellcolor{emerald5}56.67  & \cellcolor{emerald5}55.37  & \cellcolor{emerald10}10.34 \\
M95 + Mpub     & \cellcolor{emerald10}59.77 & \cellcolor{emerald10}53.01 & \cellcolor{emerald5}19.51  & \cellcolor{emerald10}62.64 & \cellcolor{emerald10}59.89 & \cellcolor{emerald5}16.51 \\
M95 + M330     & \cellcolor{emerald10}60.17 & \cellcolor{emerald10}59.09 & \cellcolor{emerald10}15.12  & \cellcolor{emerald10}63.38 & \cellcolor{emerald10}62.97 & \cellcolor{emerald10}11.09 \\
M95 + Mv0      & \cellcolor{emerald5}57.39  & \cellcolor{emerald5}51.06  & \cellcolor{emerald5}19.84  & \cellcolor{emerald10}59.96 & \cellcolor{emerald10}58.64 & \cellcolor{emerald10}13.81 \\
Mpub + M330    & \cellcolor{emerald10}63.20 & \cellcolor{emerald10}56.66 & \cellcolor{emerald10}14.60  & \cellcolor{emerald10}66.33 & \cellcolor{emerald10}64.85 & \cellcolor{emerald10}9.57  \\
Mpub + Mv0     & \cellcolor{emerald5}57.65  & \cellcolor{emerald5}49.43  & \cellcolor{emerald5}18.30  & \cellcolor{emerald10}59.29 & \cellcolor{emerald10}56.73 & \cellcolor{emerald10}13.38 \\
M330 + Mv0     & \cellcolor{emerald10}61.06 & \cellcolor{emerald10}54.20 & \cellcolor{emerald10}16.02  & \cellcolor{emerald10}64.34 & \cellcolor{emerald10}63.08 & \cellcolor{emerald10}9.38  \\
\bottomrule
\end{tabular}
\end{adjustbox}
\caption{Evalution Scores for various FM combinations}
\label{tab:complete_performance}
\end{table}

\subsection{Experimental Results}
Although the dataset considered in our study lacks musical content, we include MFMs in our experiments, believing that their pre-training on musical data might provide them strength to implicitly capture rhythmic patterns in singing voices, and thus might benefit SVDSA. Table \ref{tab:model_performancesingle} presents the evaluation results of downstream models trained with various individual FMs for SVDSA. We use accuracy, F1 score, and EER (equal error rate) as the evaluation metrics for the evaluation of the models.  Accuracy is a preferred metric for source attribution as shown by previous research for speech deepfake source attribution \cite{klein24_interspeech}. We have also used EER as it is the preferred metric for SVDD  as well as various types of audio deepfake detection \cite{zang2024singfake, chetia-phukan-etal-2024-heterogeneity}. For EER, we report the average one-vs-all scores. The results demonstrate that MMFMs consistently outperformed both SFMs and MFMs, thus proving \textit{our hypothesis that MMFMs will be the most effective for SVDSA due to their cross-modality pre-training them enables them to capture source-specific traits—such as timbre, pitch variations, and synthesis artifacts—by learning rich, complementary representations from diverse acoustic and contextual cues.} Among the MMFMs reported the topmost performance with both FCN and CNN downstreams. Overall, the CNN models shows better performance than FCN models with most of the FMs. After MMFMs, the multilingual SFMs shows the second top performance and their results can be result of their exposure to multilingual data during their pre-training as the dataset used in our study consists of singing voices in Chinese and Japanese. Further, we see that monolingual SFMs (WavLM, Unispeech-SAT, Wav2vec2) shows lower performance than its multilingual counterparts, this can be attributed to the linguistic difference between the pre-training of the monolingual SFMs and the downstream data distribution. MFMs reported the lowest performance amongst different FMs. This shows their ineffectiveness in capturing source specific characteristics. Further, a surprising observation is the performance of x-vector, a SFM trained for speaker recognition as it is a much smaller SFM compared to other SFMs. It shows better performance than both monolingual SFMs and MFMs and this performance can be due to its speaker recognition pre-training which equips it with a stronger ability to distinguish source-specific traits. We also plot the t-SNE visualization of FMs raw representations in Figure \ref{fig:tsne}. Through these plots, we observe better clustering across the source classes in MMFMs and providing support to our hypothesis and obtained experimental results. \newline
Table \ref{tab:complete_performance} presents the evaluation scores for modeling with various combinations of SFMs. We use concatenation-based fusion as baseline technique. We keep the modeling the same as \textbf{\texttt{COFFE}} except the CD loss. We also keep the training details same as \textbf{\texttt{COFFE}} for fair comparison. We observe that fusion of FMs through \textbf{\texttt{COFFE}} generally shows better performance than baseline concatenation-based fusion. With \textbf{\texttt{COFFE}}, through the fusion of MMFMs LB and IB, we report the topmost performance across all the combinations of FMs as well as individual FMs. This shows that combination of MMFMs further brings out the complementary behavior among them. These results show the efficacy of \textbf{\texttt{COFFE}} for effective fusion of FMs for improved SVDSA. Additionally, the confusion matrices in Figure \ref{fig:CM} further illustrate the improvement in classification accuracy when using \textbf{\texttt{COFFE}} with the fusion of LB and IB compared to individual LB with CNN. The results presented in our paper will act as a benchmark for future studies exploring SVDSA. 

\section{Conclusion}
In this study, we introduce the task of SVDSA and demonstrate that MMFMs are the most effective for SVDSA. MMFMs such as IB and LB, excel in capturing source-specific traits like timbre, pitch manipulation, and synthesis artifacts due to their cross-modal pretraining. Building on these insights, we propose \texttt{\textbf{COFFE}}, a novel fusion framework that leverages Chernoff Distance as a loss function to enhance the integration of FMs. \texttt{\textbf{COFFE}} through the fusion of MMFMs achieves superior performance compared to individual FMs and baseline fusion approaches, establishing a strong baseline for future research in SVDSA. Our work also calls upon researchers to work upon our established benchmarks for further performance improvement of SVDSA.

\bibliographystyle{IEEEtran}
\bibliography{main}

\end{document}